\documentstyle[12pt]{article}
\input tcilatex
\QQQ{Language}{
American English
}

\begin{document}

\begin{center}
DEDUCTION OF THE QUANTUM NUMBERS OF LOW-LYING STATES OF 4-NUCLEON SYSTEMS
BASED ON SYMMETRY

\hspace{1.0in}

C.G.Bao

Department of physics, Zhongshan University, Guangzhou, China

\hspace{1.0in}

\hspace{1.0in}

\hspace{1.0in}

Abstract

The inherent nodal structures of the wavefunctions of 4-nucleon systems have
been investigated. The existence of two groups of low-lying states with
specific quantum numbers dominated by total orbital angular momentum L=0 and
L=1, respectively, has been deduced. The understanding of the inherent nodal
structure is found to be crucial to a systematic understanding of the
spectrum.
\end{center}

\hspace{1.0in}

\hspace{1.0in}



(no figures and four tables)

\hspace{1.0in}

\hspace{1.0in}

E-MAIL ADDRESS: stsbcg@zsu.edu.cn

\newpage
In the investigation of microstructures the physicists have paid great
attention to the role of symmetry. A number of laws (or constraints)
governing the basic physical processes have been obtained [1-4]. However, it
is well known that the feature of quantum states depends on the distribution
of wavefunctions in the coordinate space, more specially on the nodal
structure of the wavefunctions. Will the nodal structure be affected by
symmetry? How does symmetry affect the nodal structure if it does? These
problems have not yet been studied systematically. Since eigenstates must be
orthogonal to each other, nodal surfaces must be introduced in higher states
so that they can be orthogonal to lower states. However there exists another
kind of nodal surfaces not arising from the requirement of orthogonality but
being imposed by the inherent symmetry of the wavefunctions (namely the
symmetry with respect to rotation, space inversion, and particle
permutation) [5]. This kind of nodal surfaces is called the inherent nodal
surface (INS). They are fixed at body-frames and can not be shifted by
adjusting dynamical parameters. In this letter, an example of a 4-nucleon
system will be used to demonstrate the origin of the INS and the effect of
them on low-lying spectra.

A systematic understanding of the energy spectra has in general not yet been
obtained for few-body systems, although a number of methods have been
developed to solve the Schr\"odinger equation to obtain precise solutions to
explain the observables. For example, it is possible to recover the ordering
of levels via a precise theoretical calculation, but it is difficult to
explain why a spectrum looks as it is. In order to understand better the
physics underlying the spectra, a qualitative study is in general necessary.
In particular, an analysis of the inherent nodal structure can uncover why
the wavefunction of a specific state is distributed in the coordinate space
in a specific way, and why a state with a specific set of quantum numbers is
lower or higher(as we shall see).

Let $\Psi _{LM}$ represent an antisymmetrized wavefunction of a 4-nucleon
system with total orbital angular momentum L, total spin S, total isospin T,
and parity $\Pi $. M is the Z-component of L. $\Psi _{LM}$ can be expanded as

\begin{equation}
\Psi _{LM}=\sum_\lambda \psi ^\lambda ,
\end{equation}

where $\lambda $ denotes the permutation symmetry of the spatial
wavefunction (a representation of the S(4) group), $\psi ^\lambda $ is
called a $\lambda $-component of $\Psi _{LM}.$ The $\lambda $ contained in $%
\Psi _{LM}$ are determined by S and T [6] as shown in Table 1.

Let i'-j'-k' be a body frame with k' being normal to ${\vec r}_{12}$ and i'
parallel to ${\vec r}_{12}$. Then

\begin{equation}
\psi ^\lambda =\sum_QD_{QM}^L(-{\gamma },-{\beta },-{\alpha }%
)\sum_if_{iQ}^\lambda {\chi }_i^{\tilde \lambda }  \label{reswidth}
\end{equation}
where $D_{QM}^L$ is the Wigner function; $\alpha ,$$\beta $ and $\gamma $
are the Euler angles that specify the orientation of the body frame; $Q$ is
the component of L along k'; $f_{iQ}^\lambda $ a spatial function of the
coordinates relative to the body frame, $\chi _i^{\tilde \lambda }$ a
spin-isospin state, ${\tilde \lambda }$ the conjugate representation of $%
\lambda $ . The subscript $i$ labels the basis functions to span the $%
\lambda $ ( ${\tilde \lambda }$ ) representation. It is noted that the $%
f_{iQ}^\lambda $ span a representation of the rotation group, inversion
group, and permutation group. This point is crucial to the following
discussion.

When the particles form a shape with a specific geometric symmetry, specific
constraints may be imposed on the wavefunction. The greater the geometric
symmetry , the stronger the constraints. For example, 

(i) If a shape contains a 2-fold axis lying along k' with the particles 1
and 2 (3 and 4) being symmetric to this axis,; i.e., ${\vec r}_{12}{\perp }{%
k^{\prime }}$ and ${\vec r}_{34}{\perp }{k^{\prime }}$ , Then a spatial
rotation about k' by 18$0^0$ is equivalent to an interchange of the
locations of particles 1 and 2, together with an interchange of 3 and 4.
Thus we have

\begin{equation}
(-1)^{Q}f^{\lambda}_{iQ} = \sum_{i^{\prime}}g^{\lambda}_{ii^{\prime}}
(p_{12}p_{34})f^{\lambda}_{i^{\prime}Q}  \label{reswidth}
\end{equation}

where the factor $(-1)^Q$ arises from the rotation, and $g_{ii^{\prime
}}^\lambda $ is a matrix element of the $\lambda $-representation (they are
known constants from the theory of the permutation group [7,8]), $p_{12\text{
}}$and $p_{34}$ denote the interchange of particles.

(ii) Additionally, if the shape has  ${\vec r}_{12}{\perp }{\vec r}_{34}{\ }$%
further, a rotation about i' by 18$0^0$ together with a spatial inversion is
equivalent to an interchange of 1 and 2. In the rotation $f_{iQ}^\lambda $
is changed to $(-1)^L$$f_{i{\bar Q}}^\lambda $ . Thus, in addition to
eq.(3), we have

\begin{equation}
\Pi (-1)^Lf_{i{\bar Q}}^\lambda =\sum_{i^{\prime }}g_{ii^{\prime }}^\lambda
(p_{12})f_{i^{\prime }Q}^\lambda \text{ .}  \label{reswidth}
\end{equation}

(iii) Additionally, if the shape has $r_{12}$ = $r_{34}$ further, the shape
is now a prolonged (or flattened) regular tetrahedron. Then a rotation about
k' by -9$0^0$ together with an inversion is equivalent to the cyclic
permutation (1423). Thus, in addition to eq. (3) and (4), we have 
\begin{equation}
\Pi i^Qf_{jQ}^\lambda =\sum_{j^{\prime }}g_{jj^{\prime }}^\lambda
[(1423)]f_{j^{\prime }Q}^\lambda \text{ .}  \label{reswidth}
\end{equation}
Eqs.(3) to (5) impose a strong constraint on the wavefunction, the $%
f_{jQ}^\lambda $ at any prolonged (or flattened) regular tetrahedron must
fulfill these equations.

(iv) Additionally, if the height length of the above prolonged tetrahedron
is equal to $\frac{r_{12}}{\sqrt{2}}$ (and $\frac{r_{34}}{\sqrt{2}})$, then
the shape is an equilateral tetrahedron (ETH). In this case each of the axes
along $\stackrel{\rightarrow }{r_i}$ (originating from the center of mass)
is a 3-fold axis. For example, a spatial rotation about the axis along $%
\stackrel{\rightarrow }{r_1}$  by 12$0^0$ is equivalent to a cyclic
permutation of particles 2, 3, and 4. Thus , in addition to eqs. (3) , (4),
and (5), we have

\begin{equation}
\sum_{Q^{\prime }}B_{QQ^{\prime }}f_{iQ^{\prime }}^\lambda =\sum_{i^{\prime
}}g_{ii^{\prime }}^\lambda [(234)]f_{i^{\prime }Q}^\lambda   \label{reswidth}
\end{equation}

where

\begin{equation}
B_{QQ^{\prime }}=\sum_{Q"}D_{Q"Q}^L(0,{\Theta },0)e^{-i\frac{2\pi }%
3Q"}D_{Q"Q^{\prime }}^L(0,{\Theta },0)  \label{reswidth}
\end{equation}

where $\Theta $ is the angle between  $\stackrel{\rightarrow }{r_1}$ and k',
cos{\ $\Theta $ } = $\sqrt{1/3}$ .

The wavefunctions at an ETH must fulfill eqs.(3) to (6).

The above constraints imposed on the shape are sufficient to specify an ETH.
For the ETH configuration, the constraints expressed by (3) to (6) are
complete, one can prove that other ''new'' constraints are equivalent
constraints. The equations (3) to (6) are homogeneous linear algebra
equations depending on L, $\Pi $, and $\lambda $. It is well known that
homoegneous equations do not always have nonzero solutions. Since the search
of nonzero solutions of linear equations is trivial, the result is directly
given in Table 2. Where, in a few cases (associated with an empty block),
there is a set of nonzero solutions $f_{iQ}^\lambda $ satisfying all these
equations; it implies that the associated $\lambda $-component $\psi
^\lambda $ (refer to eq.(2)) is nonzero at the ETH configurations and, we
may say, this $\lambda $-component is ETH-accessible. In other cases
(associated with a block with a $\times $ ) , there are no nonzero
solutions, all the $f_{iQ}^\lambda $ must be zero at ETH configurations
irrespective to the size and orientation of the ETH. In such cases, an INS
appears and the $\lambda $-component is ETH-inaccessible. This example shows
the origin of the INS.

The INS\ existing at the ETH may extend beyond the ETH. For example, when
nonzero solutions of (3) to (6) can not be found and nonzero solutions of
only (3) to (5) also can not be found, the INS will extend from the ETH to
the prolonged (or flattened) regular tetrahedron. Since an ETH\ has many
possibilities to deform (e.g., the height from a vertex to its opposite base
becomes longer or shorter), the INS\ at the ETH has many possibilities to
extend to its neighborhood. Thus, there may be a source at the ETH, where
the INS emerges and extends to ite neighborhood. In other words, a
wavefunction may have an inherent nodal structure in the domain surrounding
the ETH. The details of the inherent structure depends on L, $\Pi $, and $%
\lambda $ of the wavefunction, but does not at all depend on any dynamical
parameters. It is emphasized that if a wavefunction can access the ETH, then
it can access the neighborhood also. Therefore, the ETH-accessible
wavefunction is inherent nodeless in the large domain surrounding the ETH.

Besides the source at the ETH, another source of INS may locate at the
squares. When the particles form a square with particles 1 and 2 (3 and 4)
at the two ends of a diagonal, we have the following equivalences.  (i) a
space inversion is equivalent to $p_{12}$$p_{34}$, (ii) when k' is normal to
the plane of the square, a rotation about k' by 18$0^0$ is equivalent to an
inversion, (iii) a rotation about i' by 18$0^0$ is equivalent to $p_{34}$,
and (iv) a rotation about k' by -9$0^0$ is equivalent to a cyclic
permutation (1324). Similar to the previous discussion, these equivalences
impose constraints on the $\lambda $-components $\psi ^\lambda $ . Thus the
accessibility of the square can be thereby identified as given also in Table
2. Similarly, when a wavefunction can access the squares, it is inherent
nodeless in the domain surrounding the squares.

There may be other sources of INS (e.g., the one locates at collinear
configurations). However, since the total potential energy is much higher in
the domains surrounding the other sources, and since we are interested only
in low-lying states, we shall neglect the other sources.

Evidently, all the low-lying states tend to contain as least as possible the
number of nodal surfaces , simply because a nodal surface would cause a
remarkable increase in energy. Hence, the most important $\lambda $%
-components for the low-lying states should be both ETH-accessible and
square-accessible, which are essentially inherent nodeless (observed in the
body frame) . They are called inherent-nodeless $\lambda $-components. It is
noted that, in the higher states, an inherent-nodeless $\lambda $-component
is allowed to contain additional nodal surfaces due to requirement of
orthogonality [5]. Furthermore, it is also allowed to contain nodal surfaces
associated with the $\alpha $$\beta $$\gamma $ degrees of freedom (i.e.,
they are permitted to have an excitation of collective rotation as we shall
see in the excited states of $^4$He ).

Now it is able to deduce the quantum numbers of low-lying states
(resonances) based on an assumption that these states are dominated by
inherent-nodeless $\lambda $-components. For a 4-nucleon system, since the
size is small, the collective energy is large if L is not small. Let us
first concentrate on the case of L$\leq $1. From Table 2 it is found that
there are three inherent-nodeless $\lambda $-components, they have the (L $%
\Pi $ $\lambda $) equal to (0,+1,$\{4\}$), (1,+1,$\{2,1,1\}$), and (1,-1,$%
\{3,1\}$) , respectively. Let us see how they compose the low-lying spectrum.

For the states of S=0 and T=0, the $\{4\}$, $\{2,2\}$, and $\{1^4\}$ spatial
symmetries are allowed (refer to Table 1). Hence, the (0,+1,\{4\})
inherent-nodeless component may be contained. Accordingly, we deduce that
there is a $J^\Pi $ = $0^{+}$ and T = 0 state dominated by this
inherent-nodeless component as listed in the second row of Table 3.

For the states of S=1 and T=0, the $\{2,1,1\}$ and $\{3,1\}$ symmetries are
allowed. The former may contain the (1,+1,\{2,1,1\}) inherent-nodeless
component, while the latter may contain the (1,-1,\{3,1\}) inherent-nodeless
component. In both cases, the L and S may be coupled to J=0,1, and 2. Thus
we deduce that there is an even-parity $J^\Pi $ = $0^{+}$, $1^{+}$, and $%
2^{+}$ multiplet, together with an odd-parity $J^\Pi $ = $0^{-}$, $1^{-}$,
and $2^{-}$ multiplet. Both multiplets have L=1, as shown in Table 3.

For the states of S=2 and T=0, only the $\{2,2\}$ symmetry is allowed, where
no inherent-nodeless components can be contained (if L $\leq $ 1). Thus we
deduce that the states of S=2 and T=0 must be higher.

Based on the inherent nodal structure, without solving the Schr\"odinger
equation and without using any dynamical model, we have deduced that there
are seven low-lying T=0 states. All the other states of T=0 should be
remarkably higher , because either they are dominated by L $\geq $ 2
components, or they do not contain inherent-nodeless $\lambda $-components.
In Table 3 the J, $\Pi $, L, S, and $\lambda $ of the above predicted states
are listed, where the L, S, and $\lambda $ are only the quantum numbers of
the dominant component. Since the 4- nucleon system is small in size, the
collective rotation energy with L=1 is large. Thus we predict that there is
a large energy gap lying between the state of L=0 (the ground state) and the
six higher states of L=1.

All the T=0 levels of $^4$He below 28.31 MeV (very close to the 2n+2p
threshold) from an R-matrix analysis [9] based on experimental data are
given in Table 4. As can be seen from this table, there is a big gap lying
between the ground state and the six higher states, just as predicted.
Moreover, all the values of $J^\Pi $ of these seven states are exactly the
same as predicted. However, it may be necessary to point out that , except
the ground state and the first excited $0^{+}$ state, other states as
resonances derived from the R-matrix analysis have not yet been commonly
recognized. Although our analysis coincides with the analysis of [9], both
analyses can not assure the existence of the resonances. The six L=1 states
definitely will be split by nuclear force, the details of the split can not
be foretold by symmetry. Besides, how the width of a resonance would be
affected by symmetry remains open.

Traditionally, the spectrum has been explained based on the shell model
[10]. However, the appearance of the $0_2^{+}$ state at 20.21 MeV is not
anticipated by the shell model. In this model the 0$_2^{+}$ would contain
two quanta (2$\hbar \Omega $) of excitation, therefore it would be much
higher than the odd parity states containing only 1$\hbar \Omega $ of
excitation. But in fact the 0$_2^{+}$ is lower than all the odd parity
states. On the other hand, in our approach the appearance of the $0_2^{+}$
is explained as an excitation of collective rotation (L increases from 0 to
1) together with a change of $\lambda $ from $\{4\}$ to $\{2,1,1\}$, and a
change of S from 0 to 1. It was shown that in the calculation based on shell
model multi-$\hbar \Omega $ configurations are absolutely necessary [11].
This fact is explicit because all the excited states are not dominated by
the $\{4\}$ component (refer to Table 3), therefore the core-excitation will
be very serious and will lead to poor convergency. In fact the weight of the
0-$\hbar \Omega $ component in the T=0 0$_2^{+}$ state is only 0.08 (refer
to eq.(13) of [11]).

In the same way we can also predict the states of T=1. For odd parity
states, the (1,-1,\{3.1\}) inherent-nodeless $\lambda $-component can be
contained in [S,T] = [0,1] and [1,1] states (refer to Table 1). In the
latter case the S and L will be coupled to J=0,1, and 2. Therefore four odd
parity T=1 states are predicted as listed in Table 3. All of them have been
found in $^4$H, $^4$He (refer to Table 4), and $^4$Li with exactly the
predicted J\ and $\Pi $ [9]. Besides, there are also a number of even parity
T=1 states containing inherent-nodeless $\lambda $-components, but they are
higher and do not appear in the low-lying spectrum. This point can not be
explained simply from symmetry, but it depends seriously on the feature of
nuclear force.

The decisive effect of the inherent nodal structures presents in all kinds
of few-body systems. For one more example, if a 4-valence-electron atom
(ion) with an inert core is concerned, we have S=0,1,and 2. In accord with
S, the $\lambda $ for the spatial function equal to $\{2,2\}$, $\{2,1,1\}$,
and $\{1^4\}$, respectively. Thus, among the three inherent-nodeless $%
\lambda $-components mentioned above, only the (1,+1,$\{2,1,1\}$) is
available (if L $\leq $ 1). Thus the ground state must not be a state of
L=0, but a $^3$$P^e$ state. This fact is confirmed by all the experimental
data of the 4-valence-electron atoms (ions) including the C, $N^{+}$ , $%
O^{++}$ , $F^{+++}$ , Si, $P^{+}$ , $S^{++}$ , Ge, etc.

When a 4-boson system is concerned, only the $\{4\}$ symmetry is involved.
Let us consider the $^{16}$O nucleus as a system of four $\alpha $%
-particles. Since this system has a larger size and a heavier mass (as
compared to the $^4$He), the states with a larger L may also appear in the
low-lying spectrum. When all the L $\leq $ 4 states are concerned, three and
only three of them, namely the $0^{+}$ , $3^{-}$ , and $4^{+}$, can access
the ETH. Let us define the internal energy of a state as $E_I$ = E - $E_{rot}
$ , where $E_{rot}$ is the rotation energy. When $E_{rot}$ is roughly
estimated via an evaluation of the moment of inertia, we find that the $E_I$
of the $0^{+}$ , $3^{-}$ , and $4^{+}$ states are very close to each other,
and they are remarkably lower than the $E_I$ of the other states. [12]

The ideas and theoretical procedures proposed in this letter can be directly
generalized to study other few-body systems. In any case, the inherent nodal
structure in the domains surrounding regular shapes (e.g., a regular
octahedron in the case of a 6-body system) should be investigated [13]. The
existence of the inherent nodal structures in microscopic few-body systems
is a great marvel of quantum mechanics. Since the INS are decisive and they
are common to different systems, similarity exists surely among these
systems. Through an investigation of the INS, different systems can be
recognized via an unified point of view.

\hspace{1.0in}

\hspace{1.0in}

ACKNOWLEDGMENT: This work is supported by the National Foundation of Natural
Science of the PRC, and by a fund from the National Educational Committee of
the PRC.

\hspace{1.0in}

REFERENCES

[1] T.D.lee and C.N.Yang, Phys. ReV. {\bf 104}, (1956) 254

[2] F.A.Kaempffer ''Concepts in Quantum Mechanics'', Academic Press, 1957

[3]J.P.Elliott and P.G.Dawber, ''Symmetry in Physics'', Vol.1 and 2,
MacMillan Press LTD, 1979

[4]W.Greiner and G.E.Brown, ''Symmetry in Quantum Mechanics'',
Springer-Verlag, 1993

[5]C.G.Bao Few-Body Systems, 13, (1992) 41; Phys. Rev. A47 (1993) 1752:
Phys. Rev. Lett.79 (1997) 3475.

[6]C.Itzykson and M.Nauenberg, Rev. Mod. Phys., 38, (1966) 95

[7]D.E.Rutherford, ''Substitutional Analysis'', Edinburgh University Press,
1948

[8]J.Q.Chen, ''Group Representation Theory for Physicists'', World
Scientific, 1989

[9]D.R.Tilley et al, Nucl Phys. A541, (1992) 1

[10]A.de Shalit and H.Feshbach, ''Theoretical Nuclear Physics'', John Wily
and Sons, Inc., 1974

[11] D.C.Zheng,B.R.Barrett, J.P.Vary, W.C.Haxton, and C.-L. Song, Phys. Rev.
C52, 2488 (1995)

[12]C.G.Bao, Chin. Phys. Lett. 14 (1997) 20

[13] C.G.Bao and Y.X.Liu, in preparation.

\newpage\ 

\begin{tabular}{|c|c|c|}
\hline
S & T & $\lambda $ \\ \hline
0 & 0 & \{4\},\{2,2\},\{1$^4$\} \\ \hline
1 & 0 & \{3,1\},\{2,1,1\} \\ \hline
2 & 0 & \{2,2\} \\ \hline
0 & 1 & \{3,1\},\{2,1,1\} \\ \hline
1 & 1 & \{3,1\},\{2,2\},\{2,1,1\},\{1$^4$\} \\ \hline
2 & 1 & \{2,1,1\} \\ \hline
\end{tabular}

Table 1

\hspace{1.0in}

\hspace{1.0in}

\begin{tabular}{|c|c|c|c|c|c|c|}
\hline
&  & \{4\} & \{3,1\} & \{2,2\} & \{2,1,1\} & \{1$^4$\} \\ \hline
0$^{+}$ & ETH &  & $\times $ & $\times $ & $\times $ & $\times $ \\ \hline
0$^{+}$ & square &  & $\times $ &  & $\times $ & $\times $ \\ \hline
0$^{-}$ & ETH & $\times $ & $\times $ & $\times $ & $\times $ &  \\ \hline
0$^{-}$ & square & $\times $ & $\times $ & $\times $ & $\times $ & $\times $
\\ \hline
1$^{+}$ & ETH & $\times $ & $\times $ & $\times $ &  & $\times $ \\ \hline
1$^{+}$ & square & $\times $ & $\times $ & $\times $ &  & $\times $ \\ \hline
1$^{-}$ & ETH & $\times $ &  & $\times $ & $\times $ & $\times $ \\ \hline
1$^{-}$ & square & $\times $ &  & $\times $ &  & $\times $ \\ \hline
\end{tabular}

Table 2

\hspace{1.0in}

\hspace{1.0in}

\begin{tabular}{|c|c|c|c|c|}
\hline
T & J$^\Pi $ & L & S & $\lambda $ \\ \hline
0 & 0$^{+}$ & 0 & 0 & \{4\} \\ \hline
0 & 0$^{-}$ & 1 & 1 & \{3,1\} \\ \hline
0 & 1$^{-}$ & 1 & 1 & \{3,1\} \\ \hline
0 & 2$^{-}$ & 1 & 1 & \{3,1\} \\ \hline
0 & 0$^{+}$ & 1 & 1 & \{2,1,1\} \\ \hline
0 & 1$^{+}$ & 1 & 1 & \{2,1,1\} \\ \hline
0 & 2$^{+}$ & 1 & 1 & \{2,1,1\} \\ \hline\hline
1 & 0$^{-}$ & 1 & 1 & \{3,1\} \\ \hline
1 & 1$^{-}$ & 1 & 1 & \{3,1\} \\ \hline
1 & 2$^{-}$ & 1 & 1 & \{3,1\} \\ \hline
1 & 1$^{-}$ & 1 & 0 & \{3,1\} \\ \hline
\end{tabular}

Table 3

\hspace{1.0in}

\hspace{1.0in}

\begin{tabular}{|ccccccc|}
\hline
T & J$^\Pi $ & MeV &  & T & J$^\Pi $ & MeV \\ 
0 & 0$^{+}$ & 0 &  & 1 & 2$^{-}$ & 23.3 \\ 
0 & 0$^{+}$ & 20.2 &  & 1 & 1$^{-}$ & 23.6 \\ 
0 & 0$^{-}$ & 21.0 &  & 1 & 0$^{-}$ & 25.3 \\ 
0 & 2$^{-}$ & 21.8 &  & 1 & 1$^{-}$ & 26.0 \\ 
0 & 1$^{-}$ & 24.3 &  &  &  &  \\ 
0 & 2$^{+}$ & 27.4 &  &  &  &  \\ 
0 & 1$^{+}$ & 28.3 &  &  &  &  \\ \hline
\end{tabular}

Table 4

\newpage

Table Captions

Table 1 The allowed representation $\lambda $ of the spatial wavefunctions. $%
\{4\}$ denotes the totally-symmetric symmetry,etc.

\hspace{1.0in}

Table 2 The accessibility of the ETH (equilateral tetrahedron) and the
square configurations to the (L$\Pi \lambda $) wavefunctions. The first
column is L$^\Pi $, the first row is $\lambda .$ An empty block implies
being accessible, a block with a $\times $ implies being inaccessible.

\hspace{1.0in}

Table 3 The predicted quantum numbers of low-lying states (L$\leq 1)$. Each
of them contains an inherent-nodeless $\lambda $-components. $\lambda $ is
the permutation symmetry of the spatial wavefunctions.

Table 4,  Low-lying spectrum of $^4$He nucleus from an R-matrix analysis of
experimental data [9].


\end{document}